\documentclass[acmsmall]{acmart}

\usepackage{xcolor}
\usepackage{pifont}
\newcommand{\cmark}{\ding{51}}%
\newcommand{\xmark}{\ding{55}}%

\AtBeginDocument{%
  \providecommand\BibTeX{{%
    \normalfont B\kern-0.5em{\scshape i\kern-0.25em b}\kern-0.8em\TeX}}}

\newcommand{\tlv}{task level}
\newcommand{\llv}{loop level}
\newcommand{\tlp}{task level parallelism}
\newcommand{\llp}{loop level parallelism}
\newcommand{\pp}{pipeline parallelism}
\newcommand{\aseeker}{AccelSeeker}
\newcommand{\aseekhpvm}{AccelSeeker-HPVM}
\newcommand{\soTa}{state-of-the-art}
\newcommand{\toolname}{\texttt{Trireme}}

\newcommand{\sgemm}{\texttt{sgemm}}
\newcommand{\gemm}{\texttt{gemm-blocked}}
\newcommand{\spmv}{\texttt{spmv}}
\newcommand{\stencil}{\texttt{stencil}}
\newcommand{\lbm}{\texttt{lbm}}
\newcommand{\mdgrid}{\texttt{md-grid}}
\newcommand{\encoder}{\texttt{audio encoder}}
\newcommand{\decoder}{\texttt{audio decoder}}
\newcommand{\edge}{\texttt{edge detection}}
\newcommand{\cava}{\texttt{cava}}
\newcommand{\slam}{\texttt{SLAM}}

\definecolor{Crimsonglory}{rgb}{0.75, 0.0, 0.2}
\definecolor{applegreen}{rgb}{0.55, 0.71, 0.0}

\newcommand{\greentick}{\textcolor{applegreen}{\cmark}}
\newcommand{\redtick}{\textcolor{Crimsonglory}{\xmark}}

\begin{document}

\title{Trireme: Exploring Hierarchical Multi-Level Parallelism for Domain Specific Hardware Acceleration}

\author{Georgios Zacharopoulos}
\email{georgios@seas.harvard.edu}
\orcid{0000-0002-6644-5200}
\affiliation{
  \institution{Harvard University}
  \streetaddress{P.O. Box 1212}
  \city{Cambridge}
  \state{MA}
  \country{USA}
  \postcode{43017-6221}
}

\author{Adel Ejjeh}
\email{aejjeh@illinois.edu}
\affiliation{%
  \institution{University of Illinois at Urbana-Champaign}
  \streetaddress{201 N Goodwin Ave}
  \city{Champaign}
  \state{IL}
  \country{USA}}
  
\author{Ying Jing}
\email{yingj4@illinois.edu}
\affiliation{%
  \institution{University of Illinois at Urbana-Champaign}
  \streetaddress{201 N Goodwin Ave}
  \city{Champaign}
  \state{IL}
  \country{USA}}  
  
\author{En-Yu Yang}
\email{enyu_yang@g.harvard.edu}
\affiliation{%
  \institution{Harvard University}
  \streetaddress{P.O. Box 1212}
  \city{Cambridge}
  \state{MA}
  \country{USA}
  \postcode{43017-6221}
}

\author{Tianyu Jia}
\email{tjia@g.harvard.edu}
\affiliation{%
  \institution{Harvard University}
  \streetaddress{P.O. Box 1212}
  \city{Cambridge}
  \state{MA}
  \country{USA}
  \postcode{43017-6221}
}

\author{Iulian Brumar}
\email{ibrumar@g.harvard.edu}
\affiliation{%
  \institution{Harvard University}
  \streetaddress{P.O. Box 1212}
  \city{Cambridge}
  \state{MA}
  \country{USA}
  \postcode{43017-6221}
}

\author{Jeremy Intan}
\email{jintan2@illinois.edu}
\affiliation{%
  \institution{University of Illinois at Urbana-Champaign}
  \streetaddress{201 N Goodwin Ave}
  \city{Champaign}
  \state{IL}
  \country{USA}}
  
\author{Muhammad Huzaifa }
\email{huzaifa2@illinois.edu}
\affiliation{%
  \institution{University of Illinois at Urbana-Champaign}
  \streetaddress{201 N Goodwin Ave}
  \city{Champaign}
  \state{IL}
  \country{USA}}
  
  \author{Sarita Adve}
\email{sadve@illinois.edu}
\affiliation{%
  \institution{University of Illinois at Urbana-Champaign}
  \streetaddress{201 N Goodwin Ave}
  \city{Champaign}
  \state{IL}
  \country{USA}}
  
\author{Vikram Adve}
\email{vadve@illinois.edu}
\affiliation{%
  \institution{University of Illinois at Urbana-Champaign}
  \streetaddress{201 N Goodwin Ave}
  \city{Champaign}
  \state{IL}
  \country{USA}}  

\author{Gu-Yeon Wei}
\email{guyeon@seas.harvard.edu}
\affiliation{%
  \institution{Harvard University}
  \streetaddress{P.O. Box 1212}
  \city{Cambridge}
  \state{MA}
  \country{USA}
  \postcode{43017-6221}
}

\author{David Brooks}
\email{dbrooks@eecs.harvard.edu}
\affiliation{%
  \institution{Harvard University}
  \streetaddress{P.O. Box 1212}
  \city{Cambridge}
  \state{MA}
  \country{USA}
  \postcode{43017-6221}
}

\renewcommand{\shortauthors}{Georgios Zacharopoulos, et al.}

\begin{abstract}
The design of heterogeneous systems that include domain specific accelerators is a challenging and time-consuming process.
While taking into account area constraints, designers must decide which parts of an application to accelerate in hardware and which to leave in software.
Moreover, applications in domains such as Extended Reality (XR) offer opportunities for various forms of parallel execution, including \llv, \tlv\ and \pp.
To assist the design process and expose every possible level of parallelism, we present \toolname, a fully automated tool-chain that explores multiple levels of parallelism and produces domain specific accelerator designs and configurations that maximize performance, given an area budget.
Experiments on demanding benchmarks from the XR domain revealed a speedup of up to $20\times$, as well as a speedup of up to $37\times$ for smaller applications, compared to software-only implementations.
\end{abstract}




\maketitle

\section{Introduction}
The breakdown of Dennard scaling \cite{esmaeilzadeh2011dark}, and the seemingly inescapable end of Moore's law~\cite{simonite2016moore}, present new challenges for computer architects striving to achieve increased performance in modern computing systems.
Heterogeneous Computing has emerged to address these issues, but the complexity of heterogeneous systems, consisting of software (SW) processors and hardware (HW) accelerators, has also increased dramatically. Hardware designers assigned with accelerating a certain application domain are required to have a deep knowledge and understanding of both the software applications and the underlying platform characteristics. Additionally, a great deal of manual effort is required to identify and extract the information that is necessary to explore various possible optimizations for every design.\par

These optimizations include exploiting application level parallelism, in the form of Instruction Level (ILP), Loop Level (LLP), Task Level (TLP) and Pipeline Parallelism (PP). 
The use of such parallelism has been limited in tools for designing hardware accelerators, in two ways. First, in the few tools \cite{margerm2018tapas, schardl2017tapir} that accommodate application level parallelism, it is limited to TLP and LLP. Second, these approaches do not usually perform Design Space Exploration (DSE) in early design stages (before implementing a particular hardware design, e.g., using an HLS tool) to explore a broad range of possible designs and combinations of different types of parallelism.
\par
A hardware design 
DSE flow for a System on Chip (SoC) with hardware accelerators, that automatically extracts and uses parallelism information, requires three main components: a) A program representation that captures and exposes various levels of parallelism in an application, and also potential data movement requiring communication or memory system demands.   b) An analysis tool that explores various HW/SW partitioning options, while taking into account not only the execution time and area, but also SoC interconnect bandwidth and communication latency. c) An integration of (a) and (b), such that (a) can provide the information that (b) requires, and (b) can use this information to build efficient performance and cost models to apply in the DSE process.
\par
Spatial \cite{koeplinger2018spatial} is a tool that performs early DSE focusing on parallelism, but it has a number of limitations.  First, Spatial aims to support hardware designers by providing a hardware-centric design language, and does not support applications written in high level languages (e.g., C,C++).
Second, it is restricted to modeling performance on FPGAs and CGRAs, and cannot be used to effectively perform DSE for SoCs.  In particular, communication latency and memory bandwidth are not taken into account during DSE.  
Finally, the parts of the computation to be accelerated need to be specified by the user and no automatic exploration of acceleration candidates takes place, which is the primary goal of our work.

To address these issues we present \toolname,\footnote{Trireme was an ancient Greek/Roman boat having three main rows of oars (similar to the three types of parallelism that we explore) and requiring parallel work to flow. A lightweight, quick boat, taking advantage of parallelism is ideal for explorations, hopefully also for early Design Space Exploration.
} an automated tool-chain that integrates the AccelSeeker \cite{ZacharopoulosNov19} and Heterogeneous Parallel Virtual Machine (HPVM) \cite{kotsifakou2018hpvm} tools. AccelSeeker offers automatic identification and selection of HW accelerators based on models of performance, and HPVM is a parallel program representation for heterogeneous systems that exposes all the major forms of parallelism (\llv, \tlv\ and \pp) 
relevant to accelerator design.
We extend \toolname{} with novel models of parallel performance evaluation (described below) to enable early DSE that accounts for various forms of parallelism.
Moreover, \toolname{} is able to account for SoC interconnect bandwidth and latency, which enables strong synergy with the explicit dataflow information captured in the HPVM parallel representation (a hierarchical dataflow graph).
The integration of the two thus offers the basis for an extensive exploration of multiple levels of parallelism, provides an early estimation of performance, and outputs HW/SW designs that maximize speedup within specific area budgets.
\par

For each type of potential parallelism that
\toolname\ extracts (LLP, TLP, PP, and combinations of them), we introduce novel models of performance (in terms of latency) and area demands (in terms of hardware resources). With the aid of these models, we carry out comprehensive early DSE that selects combinations of parallel accelerator designs with increasing area budgets.
Additionally, we study a variety of architectural configurations of target 
SoCs to distinguish the impact of every type of parallelism in accordance with the characteristics and complexity of novel benchmarks from the Extended Reality (XR) domain.\footnote{Extended Reality combines Augmented, Virtual and Mixed Reality.}
\toolname\ achieves speedups of up to 20$\times$ for complex XR application components (e.g., audio decoder) and up to 37$\times$ for single-kernel applications (e.g., gemm-blocked).

Our contributions are as follows:
\vspace{-0.1cm}
\begin{itemize}
\item We present \toolname, a fully-automated tool integrating HPVM \cite{kotsifakou2018hpvm} and AccelSeeker \cite{ZacharopoulosNov19}, that offers
identification, estimation of performance and selection of hardware accelerators that exploit \tlv, \llv, and \pp{} (Section~\ref{sec:trireme}). 
\item We introduce novel models for estimating performance and resource demands (area) for \tlv, \llv, and \pp{} (Section~\ref{sec:models}).
\item 
We demonstrate \toolname's HW/SW partitioning choices while sweeping area budgets and varying the configuration of memory latency and accelerator invocation overhead, thereby covering a wide range of possible
designer scenarios (Sections~\ref{sec:setup} and \ref{sec:results}).
\item We evaluate our tool
using a broad spectrum of applications, spanning from smaller, single-kernel applications, to complex and demanding state-of-the-art application components from the XR domain (derived from a recently released XR testbed \cite{huzaifa2020exploring}) (Section~\ref{sec:results}).
\end{itemize}

\section{Background}\label{sec:background}
 
\toolname\ performs extensive early DSE of potential parallelism possibilities for HW acceleration, in comparison to tools such as TAPAS \cite{margerm2018tapas} and Peruse \cite{peruse} that offer limited or late DSE. Furthermore, our tool explores a number of different platform configurations, with respect to memory latency and overhead due to the invocation of the accelerators, that can drastically affect the performance of a HW/SW design.\par

Achieving such a thorough and early DSE, while investigating the different parallelism opportunities, is a \emph{challenging endeavor} because it requires both: a) automatic extraction of any parallelism-related information from the applications to be accelerated and b) automatic identification and early evaluation of potential accelerators. HPVM and \aseeker, both developed within the LLVM \cite{LattnerMar04} infrastructure, support the former and the latter requirements respectively, and hence, serve as the basis of the \toolname\ tool-chain.

\aseeker\ is a tool that performs automatic identification and selection of hardware accelerators, and HPVM is a parallel program representation for heterogeneous systems. \toolname\ uses components of \aseeker\ to perform an initial estimation of performance and an estimation of area requirements. We use HPVM to analyze the applications and collect required information regarding the three types of parallelism (TLP, LLP, PP) that we can exploit. In the following sections, we provide detailed backgrounds of both tools.\par

\subsection{AccelSeeker}\label{sec:background:accelseeker}

AccelSeeker is an LLVM-based tool, comprised of analysis passes, that analyzes applications represented by the LLVM Intermediate Representation (IR). It can be used in the early stages of the HW/SW partitioning process and can reveal the most promising parts of an application for HW acceleration.
The tool has three main phases: a) Candidate Identification for HW acceleration, b) Performance and Area Estimation and c) Selection of Candidates for acceleration that maximize speedup under a user-defined area constraint.\par

\begin{figure}[h]
\centering
\includegraphics[width=0.5\linewidth]{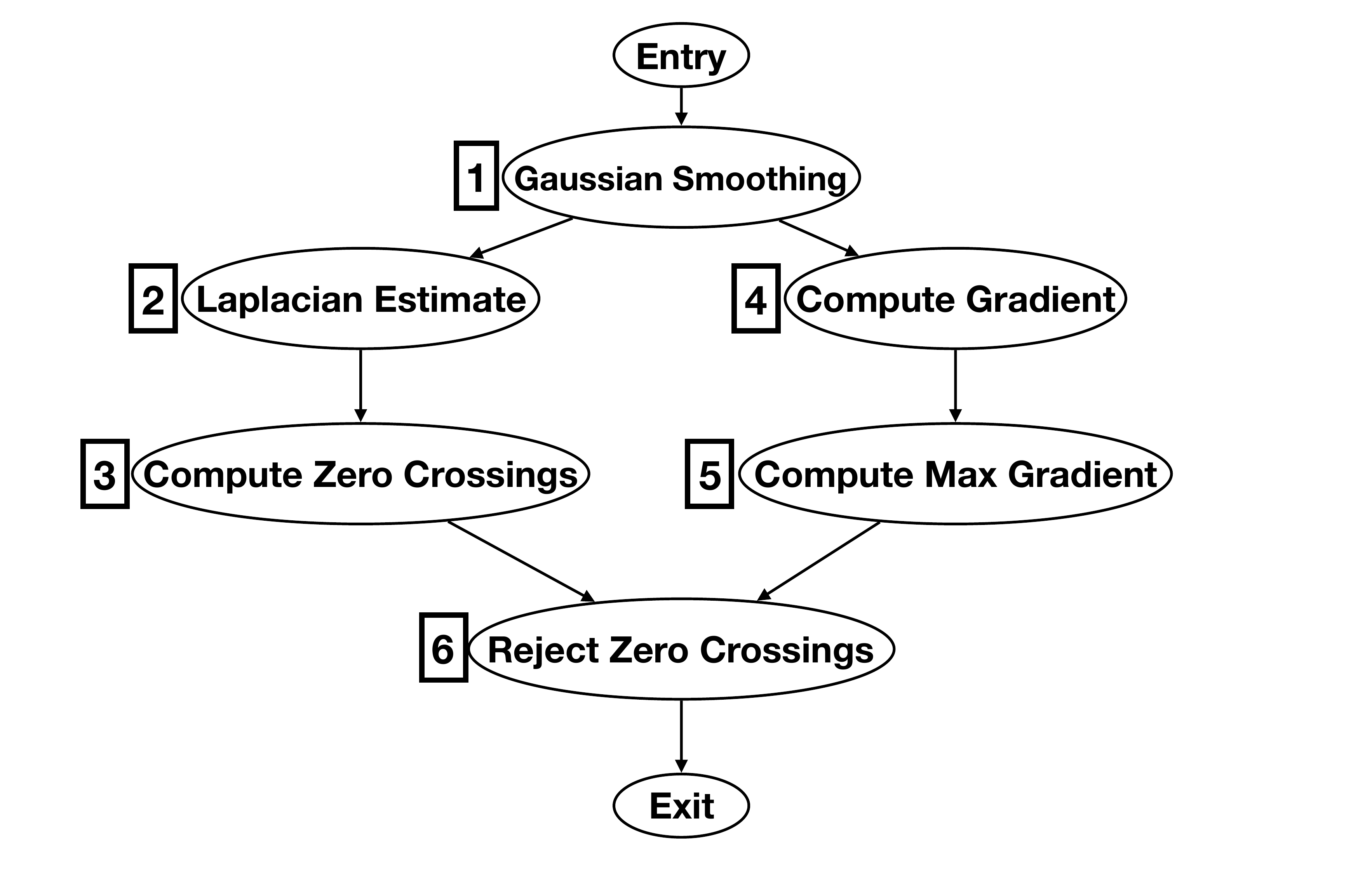}
\caption{Data Flow Graph for "edge detection".}
\label{fig:dfg-edge}
\end{figure}

\emph{Candidate Identification.} The granularity of the candidates for acceleration is defined as that of a subgraph of the call-graph of an application that satisfies two properties: It has a root and there are no outgoing edges. Effectively this translates to a candidate that is a function/task, whose calls to other functions included in it (if any) are part of its computation as a potential HW accelerator. As an example, in Figure \ref{fig:dfg-edge}, every one of edge detections's Data Flow Graph (DFG) nodes, which corresponds to a function in the call graph, can be a candidate for acceleration.
\par

\emph{Performance and Area Estimation.} \aseeker\ uses models that estimate speedup (merit) and area usage (cost). Through LLVM static analysis and dynamic profiling \cite{ZacharopoulosMar17}, these models assess software and hardware latency, area, and I/O data transfer requirements for every identified candidate. A default Zynq Programmable System-on-Chip target platform is assumed for the architectural characterization, though it can be configured to adapt to different platforms. The HW accelerators are designed as loosely coupled --- their implementations exploiting ILP within the boundaries of a Basic Block (BB). This type of accelerator, exploiting parallelism within the BB granularity, will be referred to as Basic Block Level Parallelism (BBLP) accelerators in Section \ref{sec:results}.

\emph{Selection of Candidates.} Having assigned a specific speedup estimation (merit) and HW resource requirement (cost) to every identified candidate, the selection phase takes place. For a given area budget (which can be varied from small to large) a subset of the initial candidate list is selected that maximizes speedup. The tool's output is the design of a heterogeneous system that distinguishes the part of the computation that stays in software from the part that is accelerated by hardware.

\subsection{HPVM}\label{sec:background:hpvm}

Heterogeneous Parallel Virtual Machine (HPVM) \cite{kotsifakou2018hpvm} is a parallel program representation for heterogeneous systems, designed to be a virtual ISA, compiler Intermediate Representation (IR) and run-time representation. 

Designed as an extension of LLVM IR\cite{LattnerMar04}, HPVM exploits all the optimization and code generation potential of LLVM, both for scalar and vector code, while adding support for parallel computation and heterogeneous systems. 
This is achieved by representing programs using a hierarchical Data Flow Graph (DFG).
An HPVM program consists of host code together with one or more DFGs.
All code suitable for acceleration is contained in the DFG nodes.
A DFG node can either contain a part of the computation (called a leaf node) or an entire \emph{nested} data flow graph. This hierarchical representation enables multiple levels of nested parallelism.
Every DFG node has a \textit{node function} associated with it, and node functions for leaf nodes contain ordinary scalar and vector LLVM IR.
Every DFG edge represents an explicit, logical 
data transfer between two nodes. Each static node in the graph can have multiple, independent dynamic instances specified as a replication factor (similar to the grid of threads for a CUDA or OpenCL kernel). Put together, this structure allows HPVM to capture \llv\ data parallelism (via the dynamic instances of a node), fine-grain data parallelism (via LLVM vector instructions within a leaf node), task parallelism between concurrent nodes (via pairs of subgraphs that are not connected by any path), and pipelined streaming parallelism (via streaming dataflow edges), all in a single parallel program representation.\par

The HPVM representation promotes optimizations such as node fusion, data mapping to local accelerator memory (e.g., GPU scratchpads), and memory tiling. So a number of transformations can be performed on the HPVM IR to optimize execution on specific target devices. The HPVM code generator traverses the DFG, translating each DFG node into code for one or more processing elements in the target system. The HPVM design is able to leverage LLVM's well-tuned back-ends, such as NVIDIA PTX, Intel AVX and X86-64. 
The HPVM run-time, invoked by the host code, interfaces with the corresponding device run-time to launch a kernel and copy needed data to and from the device.

\section{Trireme}\label{sec:trireme}

\begin{figure*}[t]
\centering
\includegraphics[width=0.8\linewidth]{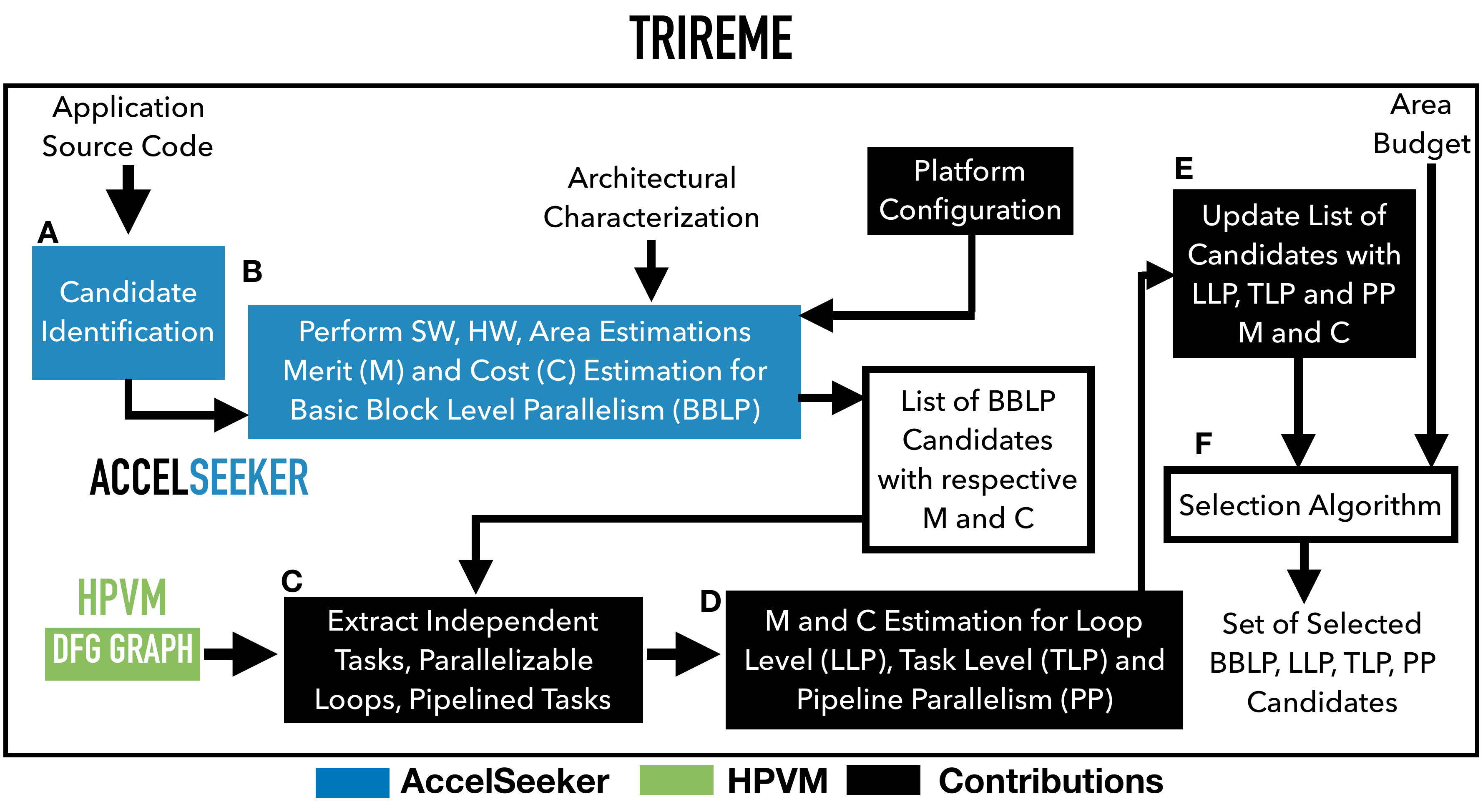}
\caption{Overview of the \toolname\ methodology.}
\label{fig:overview}
\end{figure*}

An overview of the entire methodology of the \toolname\ tool-chain is depicted in Figure \ref{fig:overview}. 
Boxes C, D and E in the figure represent new components developed for this work, while the other boxes represent existing \aseeker{} and HPVM components.
The source code (C,C++) of every application is used as input and, with the aid of \aseeker, 
we analyze its IR to identify candidates for acceleration (Box A). Next we estimate the SW and HW latency, area and the amount of data required for every identified candidate. Their potential performance gain (speedup) is estimated and attached to them as \emph{Merit}, as well as the \emph{Cost} required in terms of HW resources (Box B).\par

\begin{figure}[t]
\centering
\includegraphics[width=0.6\linewidth]{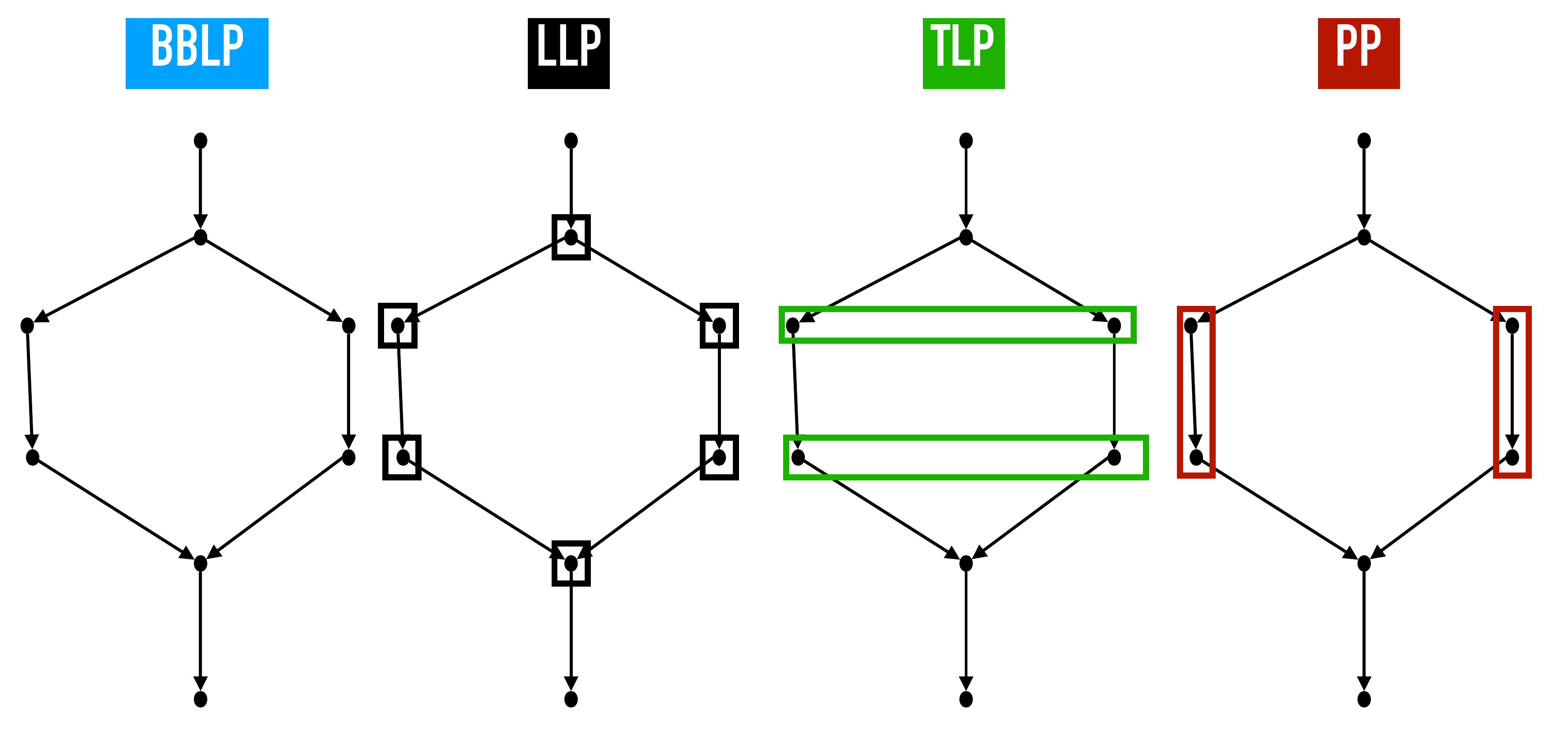}
\caption{DFG of edge detection depicting Basic Block level parallelism (BBLP) acceleration candidates, 
\llv\ (LLP), \tlv\ (TLP) and \pp\ (PP) opportunities.}
\label{fig:dfg-parallel}
\end{figure}

The list of candidates and the DFG of the application, generated by HPVM, are then passed as input to a tool that extracts all necessary information regarding potential parallel execution, as detailed in Subsection \ref{sec:ASeekHPVM}
(Box C).
With the aid of novel models for \llv\ (LLP), \tlv\ (TLP) and \pp\ (PP), described in the following sections, we estimate potential speedup (Merit) and area (Cost), including through combination of parallel approaches wherever applicable, i.e., \tlv$+$\llp\ (TLP-LLP) and \pp$+$\tlp\ (PP-TLP) 
(Box D). 
Figure~\ref{fig:dfg-parallel} shows the DFG of the edge detection benchmark and its respective parallelism opportunities.
\par

We update the list of accelerators with the newly formed candidates for acceleration that can exploit any (or all) of the three extracted types of parallelism (LLP, TLP, PP), and combinations of them 
(Box E). Finally, a selection algorithm provides the HW/SW design that maximizes the potential speedup within a given area budget (Box F).
\par

\subsection{AccelSeeker-HPVM Integration}
\label{sec:ASeekHPVM}

We have integrated \aseeker\ and HPVM to exploit any parallelism information that can be provided by the latter and guide the selection process. 
In particular, we developed a C++ tool within the HPVM infrastructure that receives a list of the most promising candidates (functions) for acceleration, as evaluated by \aseeker, along with their corresponding estimated software ($T_s$), and hardware ($T_h$) execution times. In addition, the HPVM bitcode file of the application being analyzed is provided as input. The tool builds the DFG of the provided application and creates a mapping between the DFG leaf nodes and the respective input functions from \aseeker, such that each input function corresponds to a leaf node. For the scope of this work, we only consider candidate functions that correspond to leaf nodes in the HPVM DFG. Any functions called within a leaf node are accounted for as part of the leaf node's analysis, and not analyzed separately.
The tool then performs a set of HPVM DFG analyses that extract the different types of parallelism, as described below.\par

First, a \emph{node-reachability} analysis is performed that queries the HPVM DFG to determine whether each of the candidate DFG nodes has a path connecting it to any of the other candidates. 
We consider nodes that belong to separate DFGs to be sequential. For every node $i$, we build a list of nodes that are parallel to it, such that any node $j$ that is found to be unreachable to/from $i$ is added to that list. The output of this analysis is the set of nodes that can run in parallel with each candidate.\par 

Second, a critical-path analysis is performed to calculate the Earliest Start Time ($EST$) and Earliest Finish Time ($EFT$) of each candidate node. Two full traversals through the DFG are performed: 
a) calculating the times 
while the entire run-time is in SW and b) calculating the times while the computation is implemented in HW. In each traversal, the $EST$, $EFT$, and Duration ($D$) of a leaf node ($N$) are calculated as follows:\par\noindent
$D(N)=T_s\ \text{or}\ T_h$ depending on the current traversal.\par\noindent
$EST(N)=MAX(EFT(Pred(N)))$ where $Pred(N)$ is the list of $N$'s predecessors in the graph.\par\noindent
$EFT(N)=EST(N)+D(N)$.\par
For cases with separate DFGs, we set $EST$ of the first node in a DFG $i$ to be the $EFT$ of the last node in the previous DFG $i-1$. 
The output of this analysis is the software and hardware $EST$s for each candidate function. 
This information is used in conjunction with the reachability analysis results at a later stage to determine \tlp\ (Section \ref{ssec:tlp}).\par

Finally, a third round of analysis detects for every candidate node whether or not it has dynamic replication. Its output is a table containing the nodes that have dynamic replication, along with the number of dimensions they are replicated on. Additionally, if the replication factors of a node are constants, those factors are included as well. This information is used at a later stage to determine \llp\ (Section \ref{ssec:llp}).

\subsection{Tool-chain Features}

\textbf{Accelerator Granularity.} 
We consider the granularity of the candidates to be within the boundaries of a function, as identified by an LLVM-based analysis. Furthermore, under the scope of our work, and in order to integrate AccelSeeker analysis with HPVM, HW accelerators correspond to leaf nodes in the HPVM DFG, as seen in the example of Figure~\ref{fig:dfg-edge}. In this instance, every (indexed) node of the DFG of \edge\ serves as a potential candidate for acceleration.\par

\textbf{Software, Hardware Latency and Area Estimation.}
We perform estimation of software and hardware latency for every identified candidate both by static analysis at the IR level and by extracting run-time profiling information. Furthermore, an estimation of LUTs and $mm^2$ is carried out in order to account for the hardware resource requirements of every accelerator. The former is estimated with \aseeker\ and its characterization of area in LUTs, by synthesizing a number of micro-benchmarks on a Zynq Programmable System-on-Chip (PSoC). The latter is retrieved by employing the Aladdin \cite{ShaoJul14} area characterization in $mm^2$. Our method, however, is not constrained to a specific platform and it can easily be adapted for different computing systems (e.g., FPGA boards, ASIC implementations, etc.).\par

\textbf{I/O Communication Estimation.}
The amount of data required by each candidate is also extracted by static analysis and by parsing its dynamic trace, when the latter is available. This data requirement is subsequently used to estimate latency due to communication between an accelerator and memory (e.g., DRAM, last level cache, etc.).\par

\textbf{Merit and Cost Estimation.}
Given the characteristics of the platform for which we are going to implement HW accelerators, we estimate potential speedup (Merit) for every acceleration candidate and the hardware resources required (Cost) to achieve that speedup. To obtain an accurate estimate, we use the AccelSeeker model for \emph{Merit}, which translates to cycles saved, and its model for \emph{Cost}, which accounts for the area budget in terms of LUTs (Section~\ref{sec:background:accelseeker}).

\textbf{Automatic Extraction of Parallelism.}
Using the tool developed for \aseekhpvm\ integration 
(\ref{sec:ASeekHPVM}) we automatically extract information about the potential for \llv, \tlv\ and \pp. This serves as input, along with \aseeker's list of candidates for (BBLP) acceleration, for the novel performance models of multiple levels of parallelism explored by \toolname. 
These models are presented in detail in Section \ref{sec:models}.

\textbf{Selection Algorithm.}
The updated list of candidates for acceleration is generated, including both the  Basic Block Level Parallelism (BBLP) accelerators from \aseeker\ and the candidates that exploit all types of parallelism explored by our tool-chain.
The selection algorithm recursively explores the subsets of the updated list of candidates, in a similar manner to the Bron-Kerbosch algorithm \cite{BronKerbosch73}. The output returned is the set
with the highest speedup (cumulative Merit) that stays within the user defined area budget (Cost).

\begin{figure}[t]
\centering
\includegraphics[width=0.65\linewidth]{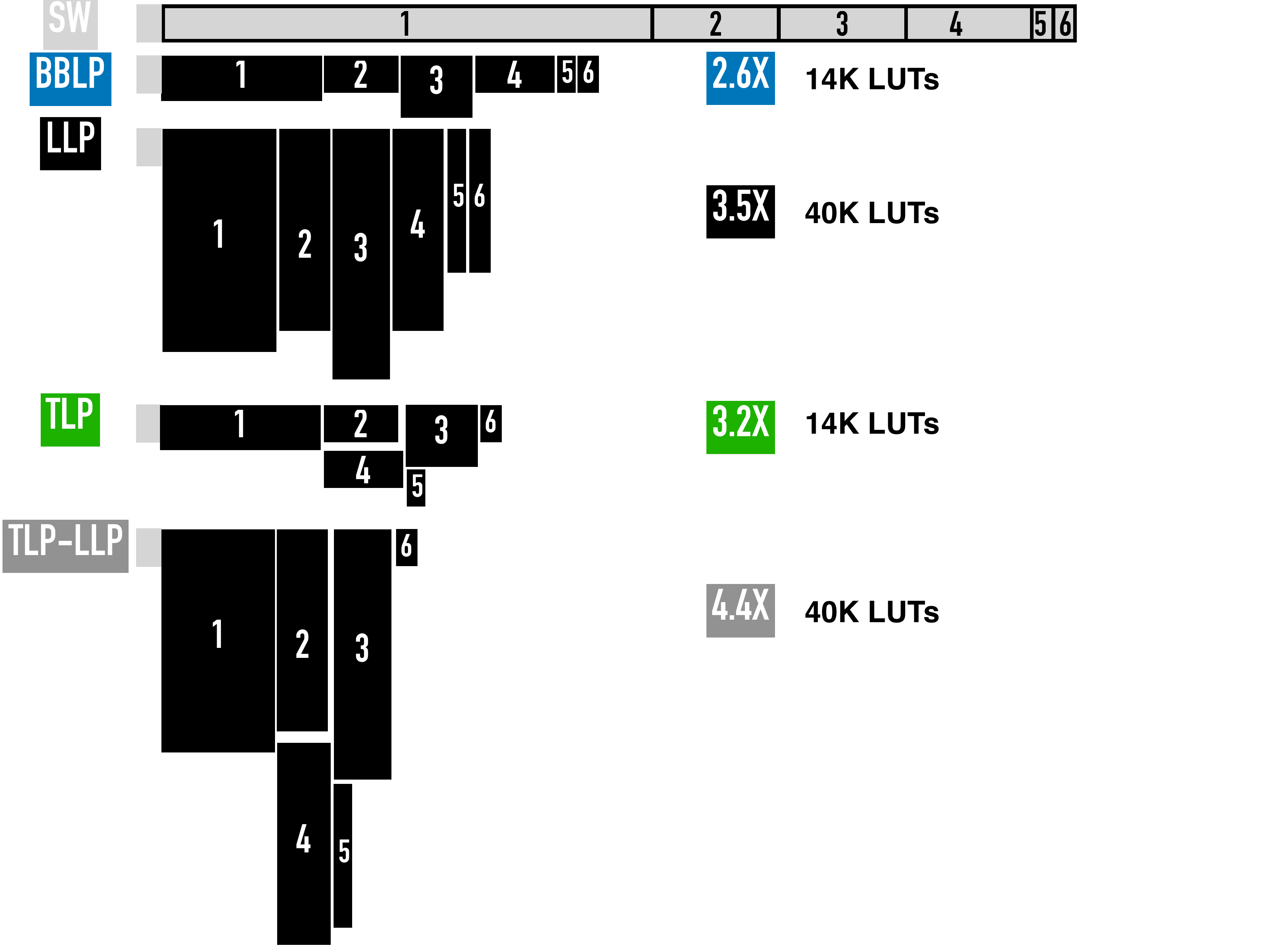}
\includegraphics[width=0.65\linewidth]{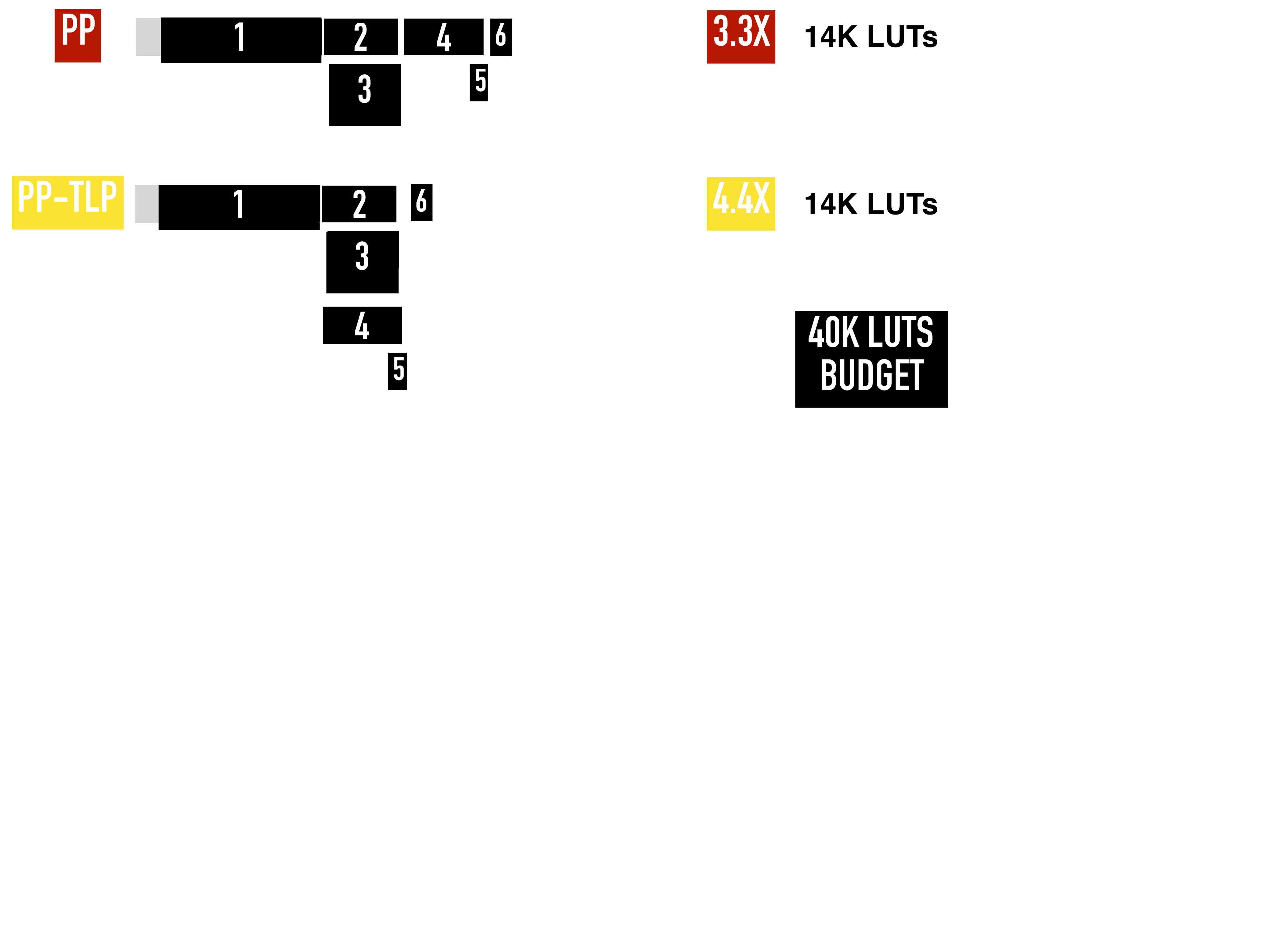}
\vspace{-3.5cm}
\caption{Designs exploiting Basic Block level (BBLP - AccelSeeker \cite{ZacharopoulosNov19}), \llv\ (LLP), \tlv\ (TLP), and \pp\ (PP) in edge detection provided a $40 \times 10^3$ LUTs area budget. The size of the black rectangles represents area usage.}
\label{fig:paralel_types}
\end{figure}

\section{Merit and Cost Models}
\label{sec:models}

As mentioned in the previous section, we introduce novel models for estimation of speedup, which we denote as \emph{Merit}, and an estimation of the area required for every HW accelerator implementation, denoted as \emph{Cost}. 
These models, inspired by the respective ones from RegionSeeker \cite{ZacharopoulosApr19} and AccelSeeker \cite{ZacharopoulosNov19}, are introduced to accommodate the estimation of \llv, \tlv\ and \pp\ extracted by HPVM. Having an early estimation of speedup and area budget needs, for every possible design exploiting any, or a combination, of these three types of parallelism can lead to better design choices and significantly less engineering effort.

\subsection{Loop Level Parallelism (LLP)} \label{ssec:llp}

With the aid of the tool described in \ref{sec:ASeekHPVM}, information regarding the DFG nodes loop-level parallelism is retrieved. As shown in the example of Figure \ref{fig:dfg-parallel}, the marked DFG nodes of edge detection are identified as nodes that    
contain a fully-parallelizable loop and, thus, are analyzed further so that multiple versions of the same functions are generated with an increasing LLP
factor. For each factor, the loop is parallelized by replicating its body, and the corresponding speedup  and cost estimates are computed. To simplify the estimation we assume an equal workload for every iteration of the loop.\par

\emph{LLP Merit and Cost Estimation.}
Let $S = \{ S_1, S_2, \ldots, S_N \}$ be a set of parallelizable loop candidates, with associated SW latency ($SW_i$), HW computation latency ($HWcomp_i$), HW communication latency ($HWcom_i$), invocation overhead ($OVHD_i$) and area cost ($A_i$). Also let LLP factor $j=1\dots,K\ |\ K=max(Loop\ Trip\ Count)$ be the factor by which we parallelize each loop.  To simplify the analysis, we assume that the loop is perfectly load-balanced, and communication latency is constant, independent of $j$.
\par
Under these simplifying assumptions, for every loop candidate $\{S_{ij}\ |\ i=1,\dots, N$, $\ j=1,\dots,K\}$,
we compute the merit 
$M(S_{ij}) = SW_i- {HWcomp_i \over j} - HWcom_i - OVHD_i$, 
and the loop candidate area cost 
$C(S_{ij}) = A_i \times j$, respectively.

As anticipated, by increasing the replication factor, better performance is achieved with the higher cost of area required. LLP, where applicable, can yield tremendous speedup benefits but at a high area budget cost, as
seen in Figure \ref{fig:paralel_types} (LLP vs software-only implementations) and discussed in greater extent in Section \ref{sec:results}.

\subsection{Task Level Parallelism (TLP)} \label{ssec:tlp}
To compute the potential speedup of a number of tasks that can be run in parallel we need first to extract all possible sets of independent candidates, i.e., all candidates that have no data flow dependencies. As depicted in the example in Figures \ref{fig:dfg-edge} and \ref{fig:dfg-parallel}, edge detection candidates indexed \{2,4\} and \{3,5\} are independent sets and can therefore be invoked in parallel. The same applies for candidates \{2,5\} and  \{3,4\}. For this analysis, we use the SW and HW estimated times, as well as the EST provided from the tool described in~\ref{sec:ASeekHPVM} as described below.\par

\emph{Merit and Cost Definition/Estimation of TLP.}
Let $S = \{ S_1, S_2, \ldots, S_N \}$ be a set of independent candidates (tasks), with associated SW latency ($SW_i$), HW computation latency ($HWcomp_i$), HW communication latency ($HWcom_i$), invocation overhead ($OVHD_i$) and area cost ($A_i$).  In the best case, all candidates in the set will be able to start execution at the same time, and the total HW latency of this set of candidates $S$ would be $MAX(S_{H_W}) = max(HWcomp_i + HWcom_i + OVHD_i)\ |\ i=1,\dots, N\ $.
\par
In practice, some candidates may have varying starting times (e.g., \{2,5\}) because of dependences on previous tasks not in the candidate set (e.g., 5 must wait for 4 to complete).  To account for these delays, we add an extra overhead based on the difference of ESTs of the nodes in the candidate set: $EST\_OVHD = max(EST_i)  - min(EST_i) | i=1,\dots, N$.  Intuitively, the overhead allows us to mark the candidate set \{2,4\} as a better candidate for acceleration compared to \{2,5\}. 
\par

We denote the merit of set $S$, by $M(S) = \sum_{i\in [1,N]}{ SW_i} - MAX(S_{H_W}) - EST\_OVHD$ and we denote the cumulative cost of set $S$ in area by $C(S) = \sum_{i\in [1,N]}A_i$.\par
Task level parallelism, in applications that have independent tasks, can offer significant speedup compared to, for instance, sequential accelerators exploiting only Basic Block level parallelism (BBLP) that require the same HW resources. Figure \ref{fig:paralel_types} provides a comparison between TLP and BBLP when accelerating the edge detection application.

\subsection{Pipeline Parallelism (PP)}

We assume that the pipeline has \(K\) stages, \(S_{1}\), \(S_{2}\), ... \(S_{K}\), and the time needed on stage \(i\) is \(T_{i}\). We also assume that the stage that requires the longest time is \(S_{j}\) (i.e., $\forall i \in \{1, 2, ..., K\}, T_{j} \geq T_{i}$). Now we will prove that the total execution time for pipeline parallelism is \(T_{total} = \sum_{i=1}^{K}{T_{i}} + {T_{j}} \times (N - 1)\), where $N$ is the number of iterations.

The first term \(\sum_{i=1}^{K}{T_{i}}\) is the time spent on the first iteration. The second term \(\max_{i}{T_{i}} \times (N - 1)\) is the timing overhead caused by the following \((N-1)\) iterations. 

\textit{Step 1.} 
Provided that the inter-stage dependencies are not considered (e.g., \(S_{2}\) cannot start before \(S_{1}\) finishes, etc.), the earliest starting time for each stage is the ending time of the same stage in the previous iteration. 

If we start the second iteration after \(T_{j}\), since $T_{j} \geq T_{i}, \forall i \in \{1, 2, ..., K\}$, the starting time of every stage in the second iteration will be no later than the ending time of the same stage in the previous iteration. In other words, there should not be any idle time. 
Thus, the ending time for the second iteration is \(T_{j}\) later than the first iteration.

For the third iteration the ending time is \(T_{j}\) later than the second iteration. Thus, based on mathematical 
induction, we can prove that at iteration \(n\), execution is completed \(T_{j}\) later than the previous iteration.

\textit{Step 2.} We prove that iteration \(n\) cannot finish at time \(t\) later than the previous iteration, if \(t < T_{j}\). 
Provided that the ending time of the second iteration is \(t\) later than the first iteration,  the ending time of stage \(S_{K}\) in the second iteration is \(t\) later than the one in the first iteration. Therefore, the starting time of stage \(S_{K}\) in the second iteration is \(t\) later than the one in the first iteration. Due to the inter-stage correlation, the ending time of stage \(S_{K-1}\) in the second iteration should be no more than time \(t\) later than the one in the first iteration.

\begin{figure}[h]
\centering
\includegraphics[width=0.6\linewidth]{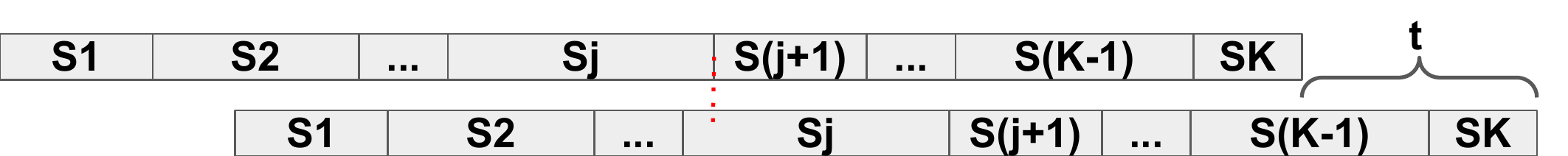}
\caption{Overlap on stage \(S_{j}\).}
\vspace{-0.3cm}
\label{fig:ppOverlap}
\end{figure}

Hence, if we trace back to \(S_{j}\), we can say that the ending time of \(S_{j}\) in the second iteration should be no more than time \(t\) later than the one in the first iteration. However, since \(t < T_{j}\), the starting time of \(S_{j}\) in the second iteration will be \(T_{j} - t\) earlier than the ending time of \(S_{j}\) in the first iteration. In other words, there will be an overlap between two consecutive iterations on stage \(S_{j}\) (Figure~\ref{fig:ppOverlap}).

\emph{Merit and Cost Definition/Estimation of PP.}
Based on the previous illustration, let $S = \{ S_1, S_2, \ldots, S_K \}$ be a set of pipelined candidates (tasks) and $N$ be the number of iterations, with associated SW latency ($SW_i$), HW computation latency ($HWcomp_i$), HW communication latency ($HWcom_i$), invocation overhead ($OVHD_i$), 
HW latency $HW_i = HWcomp_i+HWcom_i+OVHD_i$ 
and area cost $(A_i)$.
We compute the HW latency, using the previous proof, as $HW_{TOTAL} = \sum_{i=1}^{K}{HW_i} + \max_{i}{HW_i} \times (N - 1) $. This formula can be applied to both a balanced pipeline and an unbalanced pipeline.
\par
We denote the merit of set S, by $M(S) = \sum_{i\in [1,K]}{ SW_i} - HW_{TOTAL} $ and we denote the cumulative cost of set S in area by $C(S) = \sum_{i\in [1,K]}A_i$.\par


\section{Experimental Setup}
\label{sec:setup}

For our experiments, we assume a heterogeneous system constituted by a single SW processor and multiple loosely coupled HW accelerators. The processor invokes the accelerators via a memory-mapped interface. DMA is used to transfer data from main memory to accelerator scratchpads and vice versa in order to store the accelerators output to main memory and be available to the SW processor. 
As AccelSeeker, used as a baseline, targets by default an FPGA SoC (Zynq UltraScale SoC), we also use FPGA SoCs in our experiments.

\textbf{Benchmarks.} We evaluated the \toolname\ tool-chain in a variety of applications, spanning from smaller, single-kernel ones, to larger and more demanding ones. The type of potential parallelism extracted from every benchmark, as expected, also varies. The kernels from Parboil \cite{stratton2012parboil} and MachSuite \cite{reagen2014machsuite} offer opportunities for \llp\ only. Medium and large size applications from the XR domain, such as 3D spatial \encoder\ from a recently released XR testbed \cite{huzaifa2020exploring} and Camera Vision Pipeline \cava \cite{yaoyuannnn}, where both \llp\ and pipelining would be feasible, and visual inertial odometry (VIO), often referred to as \slam, where 70\% of its run-time is evaluated and  \llv\ and \tlp\ opportunities are present. Larger and more complex applications, where all types of parallelism can be retrieved (as well as combinations of them), are also rigorously evaluated. These include 3D spatial \decoder\ (XR domain) from the XR testbed \cite{huzaifa2020exploring} and \edge, a six stage image processing pipeline used in \cite{kotsifakou2018hpvm}. 

\textbf{Parallelism Strategies.} We evaluate and compare the following parallelism strategies for HW acceleration:\\
\textbf{a) Basic Block Level Parallelism (BBLP)}. Function (Task) accelerators that exploit Instruction Level Parallelism within a Basic Block. It corresponds to the accelerators selected by AccelSeeker \cite{ZacharopoulosNov19}.\\
\textbf{b) Loop Level Parallelism (LLP)}. Replication and parallel execution of fully parallelizable loops, represented in HPVM as leaf nodes with multiple dynamic instances.\\
\textbf{c) Task Level Parallelism (TLP)}. Sets of two or more tasks (HPVM leaf nodes) that have no data flow dependencies between them (i.e., no path in the HPVM dataflow graph connecting any pair of nodes in the set) and can therefore all run in parallel with each other.\\
\textbf{d) Pipeline Parallelism (PP)}. Sequences of HPVM nodes (tasks) connected by streaming dataflow edges, and therefore can be pipelined.\\
\textbf{e) Task and Loop Level Parallelism (TLP-LLP)}.
 Sets of tasks that can be either executed as parallelizable loops or run as parallel tasks or both. The final design may have any of these forms of parallelism applied.\\ 
\textbf{f) Pipeline and Task Level Parallelism (PP-TLP)}. Sets of pipelined tasks that can also be run in parallel.

\textbf{Validation.} For the validation of our models we evaluated HW acceleration with Aladdin \cite{ShaoJul14} HW accelerator simulator. The run-time of the non-accelerated part was measured using gem5 \cite{BinkertFeb11}. The processor modelled is an ARMv8-A processor of issue width of 1, having an atomic model, in-order execution and clocked at 100 MHz. Additionally, we used Catapult HLS\cite{CatapultHLS} to synthesize the HW accelerators for further validation.


\section{Experimental Results}
\label{sec:results}

In the following subsections, we showcase the speedup achieved from the hierarchical multi-level parallelism strategies explored by our tool-chain.
We group the results by different types of parallelism exploited by \toolname{}.
First, the performance benefits in single-kernel applications that solely exploit LLP are presented. Then, we investigate XR applications with pipelines (\texttt{audio encoder, cava}) and independent tasks (\slam), where both LLP/PP or LLP/TLP can be applied. Finally, we study larger ones (\texttt{audio decoder, edge detection}), where LLP/TLP/PP and combinations of them can be used, such as TLP-LLP and PP-TLP, as described in the previous section. We evaluate the above against SW-only implementations, and against \soTa{} AccelSeeker. As such, we target FPGA SoCs in all our experiments. \par

We validate the designs selected by our tool, given increasing area constraints, first using Aladdin \cite{ShaoJul14} (for the latency of HW accelerators) and gem5 \cite{BinkertFeb11} (for the software latency), and second using Catapult HLS for real hardware measurements. Finally, we study the effects of varying the bandwidth of data transfers between host and accelerator, and the overhead of accelerator invocation, on the \texttt{audio decoder} and \texttt{edge detection} benchmarks.

\subsection{Loop Level Parallelism}
\label{sec:res_llp}

\begin{figure*}[t]
\centering
\includegraphics[width=0.32\textwidth]{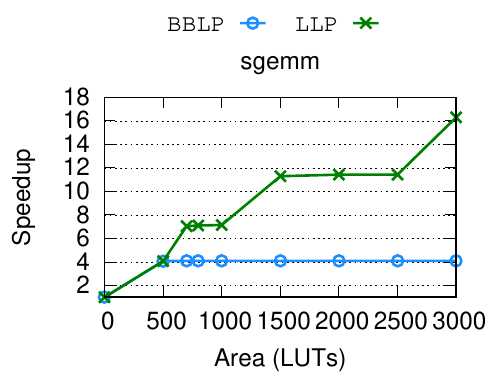}
\includegraphics[width=0.32\textwidth]{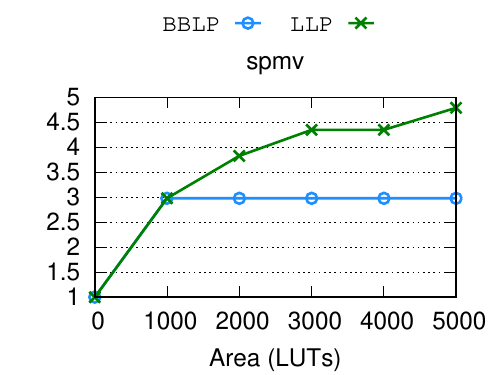}
\includegraphics[width=0.32\textwidth]{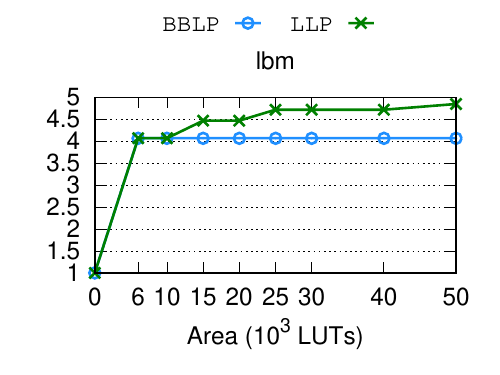}
\hspace{-1.5cm}
\includegraphics[width=0.32\textwidth]{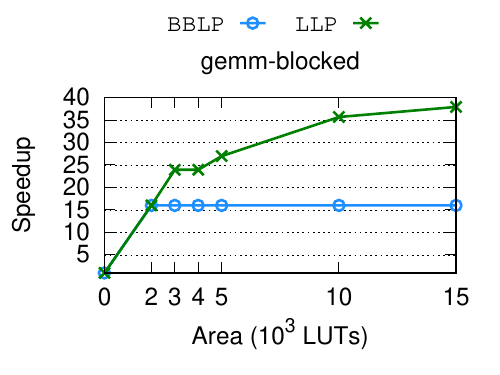}
\includegraphics[width=0.32\textwidth]{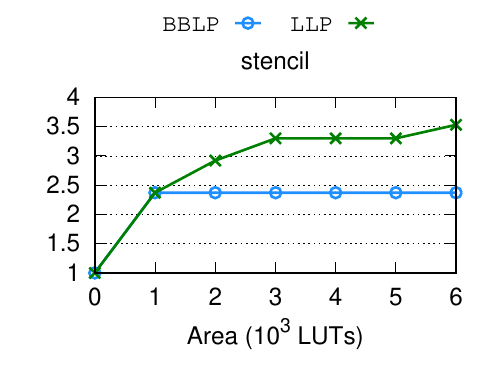}
\includegraphics[width=0.32\textwidth]{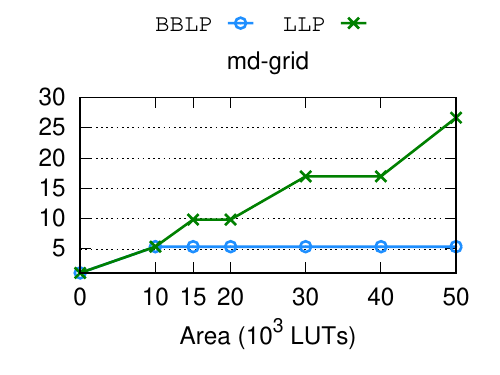}
\vspace{-0.2cm}
\caption{Speedup obtained for 
applications from Parboil \cite{stratton2012parboil} and MachSuite \cite{reagen2014machsuite} benchmark suites, varying the area budget constraint. We evaluate AccelSeeker \cite{ZacharopoulosNov19} (BBLP) and LLP, while the baseline is a SW-only implementation.}
\vspace{-0.2cm}
\label{fig:llp_results}
\end{figure*}

\toolname, extracting information exposed by HPVM, identifies the application kernels that contain a fully parallelizable loop or loop nest. Subsequently, the Merit/Cost estimation models for \llp, as described in Section \ref{sec:models}, are used to estimate the speedup and hardware resource utilization for varying LLP factor. Figure \ref{fig:llp_results} shows the speedup obtained on six benchmarks from Parboil (\texttt{sgemm, lbm, spmv}) and MachSuite (\texttt{gemm-blocked, md-grid, stencil}), compared to a SW-only baseline.\par

All applications benefit significantly from replicating their loop-bodies and running them in parallel, and the parallelism enables the designs to take advantage of larger area resources to achieve greater speedups than is possible without \llp.  For an area budget of $3\times 10^3$ LUTs, \sgemm\  and\ \gemm\ reach a 16$\times$ and 25$\times$ speedup respectively, compared to the baseline, and a 3$\times$ and \textasciitilde2$\times$ speedup compared to BBLP, which corresponds to \soTa\ \aseeker\ selections.\par

Kernels such as \spmv\ and \stencil{} realize a 4.7$\times$ and 3.4$\times$ speedup compared to a SW-only implementation respectively, for a budget of $5\times 10^3$ LUTs, whereas \lbm\,
having a smaller loop body, i.e., fewer instructions and less computation time within the loop body compared to the previous ones,
has little benefit from extra area resources and LLP. Finally, \mdgrid\ requires more area compared to the previous kernels and, having a large potential for \llp, reaches a 27$\times$ speedup compared to the SW baseline and 5.4$\times$ compared to \soTa\ BBLP accelerators.  Overall, \toolname{} is able in many cases to achieve substantial performance improvements for given hardware resources by exploiting \llp{} alone.

\subsection{Loop vs. Pipeline and Loop vs. Task Parallelism}
\label{sec:res_llp_pp}

Richer applications, such as components from the XR testbed \cite{huzaifa2020exploring}, contain a variety of opportunities to exploit parallelism. For \encoder\ and \cava, in addition to parallelizable loops, the DFG nodes can also be pipelined. For \slam, apart from LLP, independent tasks are present as well. \toolname\ automatically generates designs exploiting this information.
\par

\begin{figure*}[t]
\centering
 \hspace{-0.5cm}
\includegraphics[width=0.33\linewidth]{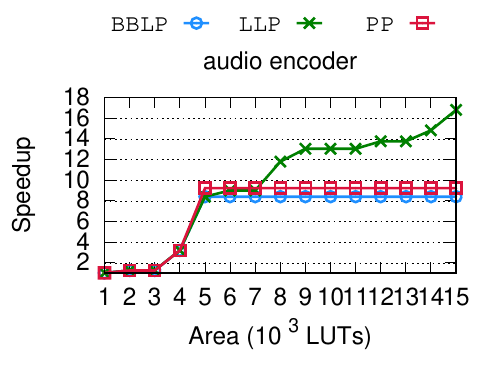}
\includegraphics[width=0.33\linewidth]{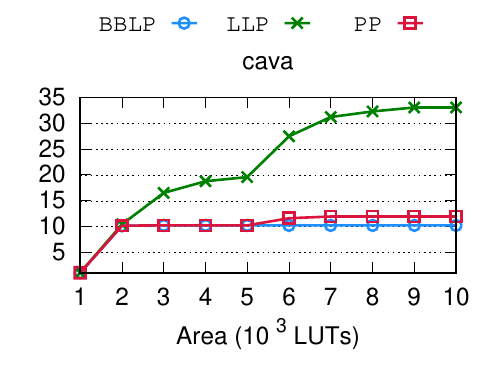}
\includegraphics[width=0.33\linewidth]{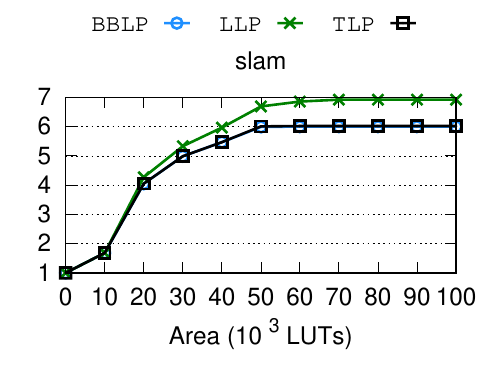}

\vspace{-0.4cm}
\caption{
Speedup obtained over the entire run-time of audio encoder \cite{huzaifa2020exploring}, cava \cite{yaoyuannnn} and  
OpenVINS algorithm for SLAM \cite{huzaifa2020exploring}, varying the area constraint. 
We evaluate AccelSeeker \cite{ZacharopoulosNov19} (BBLP),
LLP, PP and TLP, while the baseline is a SW-only implementation.
}
\vspace{-0.2cm}
\label{fig:enc-cava}
\end{figure*}

Figure \ref{fig:enc-cava} shows the speedup obtained from applying LLP and PP on \encoder\ and \cava, for a number of increasing area budgets. For a budget of $5 \times 10^3$ LUTs \encoder\ achieves an 8$\times$ (for LLP) and 9$\times$ (for TLP) speedup compared to SW-only baseline, as the entire pipeline fits the budget. Additionally, a slight improvement over BBLP (AccelSeeker selection) is achieved. Nonetheless, more area is required to parallelize the loops within the selected accelerators, which is evident by the increasing trend line for LLP. \par

For the same area budget in \cava, the pipeline does not fit. Thus, the speedup gain for PP is the same as for BBLP (10$\times$ over the baseline). LLP on the other hand benefits from loop parallelization and achieves a 20$\times$ speedup.\par

For larger budgets, we can observe significant benefits in speedup for LLP, both in \encoder\ and \cava. With $15 \times 10^3$ LUTs \encoder\ achieves a \textasciitilde17$\times$ speedup compared to baseline, and with $10 \times 10^3$ LUTs \cava\ attains a 33$\times$ speedup.
These are respectively about 2$\times$ and 3$\times$ the speedup achieved with BBLP alone.

Figure \ref{fig:enc-cava} shows that \slam\ benefits from LLP, reaching up to 7$\times$ speedup, as the area budget allows for more \llp. On the other hand, since only two tasks --- with small latency relative to the total run-time --- can be parallelized, TLP offers no performance gain.\par

For \encoder\ and \cava, PP produces little improvement in performance. This is due to the unbalanced pipelines in these workloads. One of the functions (DFG nodes) in each application dominates the computation time, therefore applying the PP strategy yields little benefit.
However, as demonstrated in the following round of experiments, this is not the case for the next two applications evaluated: \decoder\ and \edge.\par

\subsection{Loop/Task/Pipeline Parallelism}
\label{sec:res_llp_pp_tlp}

\begin{figure*}[t]
    \centering
      \hspace{-1cm}
    \includegraphics[width=0.45\linewidth]{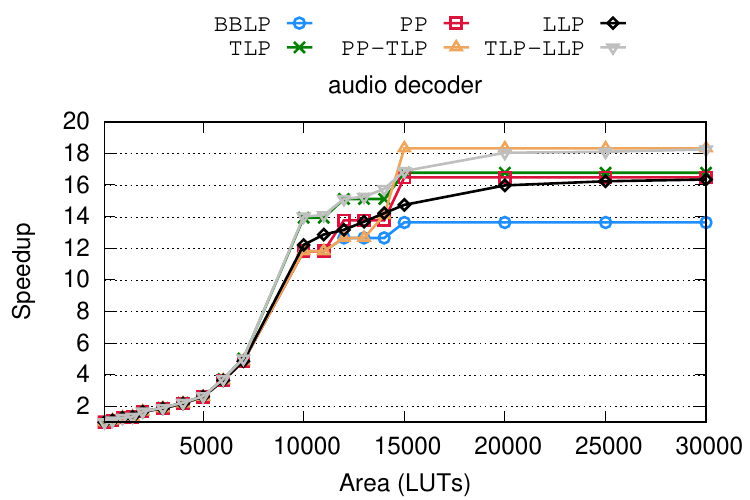}
    \includegraphics[width=0.45\linewidth]{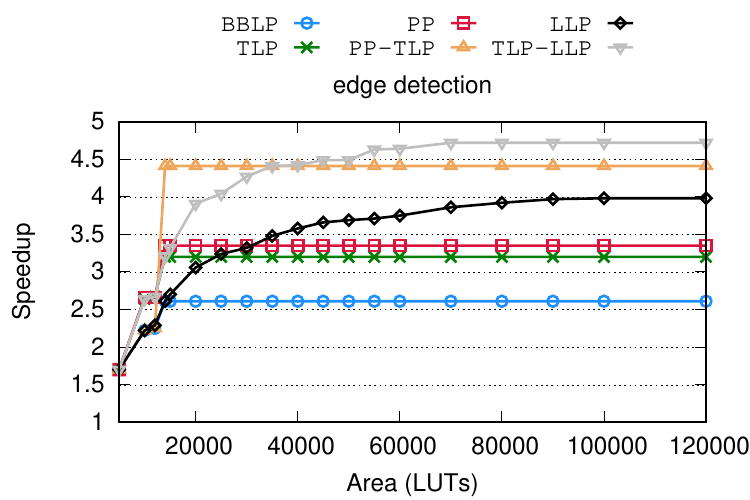}
    \vspace{-0.2cm}
    \caption{
    Speedup over the entire runtime for different versions of audio decoder (left) and edge detection (right), varying the area constraint. The baseline is a SW-only implementation.
    }
    \label{fig:audio_decoder}
\end{figure*}

In the previous subsection we encountered applications that could only exploit LLP and PP, whereas \decoder, a \soTa\ XR application component, and \edge, a six-stage image processing pipeline, can offer LLP, TLP, PP, as well as combinations of them. Such applications are ideal candidates to employ \toolname\ and unlock their full parallelism potential. Figure \ref{fig:audio_decoder} presents the speedup achieved by multiple levels of parallelism explored by our tool-chain, for increasing area budgets.\par

On \decoder, Figure \ref{fig:audio_decoder} (left) and Table \ref{tab:decoder}, for an area budget of $12 \times 10^3$ LUTs, LLP and PP reach a 13.2$\times$ and 13.7$\times$ speedup respectively, compared to a SW-only baseline. This budget is enough to fit one of the two \decoder\ pipelines, and since the workloads are fairly balanced, we see the benefit obtained from this strategy. TLP and TLP-LLP achieve the same 15.1$\times$ speedup, as not enough area is available to benefit from parallelizing the loops, while the selected independent tasks are accelerated in parallel.\par

Increasing the budget to $14 \times 10^3$ LUTs, almost equivalent to Xilinx Artix Z-7007S PSoC \cite{ZynqMar17}, we can see that LLP and TLP-LLP are making use of the larger area and increase their respective speedups to 14.21$\times$ and 15.74$\times$. Conversely, BBLP, TLP and PP extract no benefit, using only 85\% of the available resources, as their potential candidate choices require more area to be selected (Table \ref{tab:decoder} - row 2). A budget of  $15 \times 10^3$ LUTs, however, accommodates all available tasks to be parallelized (TLP-16.7$\times$), as well as the pipelines (PP-16.5$\times$), including the possibility to parallelize the independent pipelines (PP-TLP-18.31$\times$), yielding the maximum possible speedup for these strategies.\par
The latter point can also be seen in the last row of Table~\ref{tab:decoder}. A larger area budget, almost equivalent to Xilinx Artix Z-7012S PSoC \cite{ZynqMar17}, allows LLP and TLP-LLP to benefit from increased parallelization of the loop bodies of their accelerators. TLP, PP and PP-TLP show no benefit from the doubling of the hardware resources as they have already reached their better-performing designs.
An interesting aspect is that PP-TLP, the strategy that achieves the best speedup, along with TLP and PP require 
fewer hardware resources to reach their maximum speedup compared to LLP and TLP-LLP, the latter achieving an almost equivalent speedup to PP-TLP but for much larger area.
Also BBLP is consistently outperformed by all parallelism strategies explored.\par
Similar trends can be seen in \edge\ while investigating its potential for parallelism (Figure \ref{fig:audio_decoder} -- right). For a $14 \times 10^3$ LUTs area budget TLP (3.2$\times$), PP (3.4$\times$) and PP-TLP (4.4$\times$) can accommodate all their respective HW/SW designs and reach their top speedups compared to the SW-only baseline. For the same budget, LLP and TLP-LLP can achieve 2.5$\times$ and 3.2$\times$ respectively, requiring more area to reach better performance. An area budget of $40 \times 10^3$ LUTs, equivalent to Artix Z-7014S PSoC, would allow for more parallelization of the loop bodies for LLP an TLP-LLP, the latter reaching an equivalent of the PP-TLP maximum speedup (4.4$\times$).\par
For even larger area budgets, such as $100 \times 10^3$ LUTs, we notice that LLP reaches a 4$\times$ speedup and TLP-LLP surpasses the highest-performing PP-TLP design by achieving 4.7$\times$ speedup compared to the baseline. This is because, unlike \decoder, all of the accelerated functions in \edge\ have parallelizable loops, which allows for increasing speedup as the area increases.

\begin{table}[h]
  \footnotesize
  \centering
  \resizebox{0.6\linewidth}{!}{
    \begin{tabular}{|l|c|c|c|c|} 
    \hline
    \textbf{Benchmark}    & \textbf{Parallelism}   &  \textbf{Area Budget} & \textbf{Area Used} & \textbf{Speedup}   \\
  & \textbf{Version} & (LUTs) & (LUTs) & \textbf{vs. SW} \\ 
   \hline \hline
    audio decoder & BBLP & 12 000 & 11916 (99\%) & 12.65 \\
             & LLP  &  & 11655 (97\%) & 13.2 \\
            & TLP    &   & 11916 (99\%) & 15.1 \\ 
            & TLP-LLP &   & 11916 (99\%) & 15.1 \\ 
            & PP    &  & 11916 (99\%) & 13.7 \\  
            & PP-TLP    &  & 11916 (99\%) & 12.65 \\ \hline
        & BBLP & 14 000 & 11916 (85\%) & 12.65 \\   
            & LLP  & \textbf{Artix Z-7007S}   & 13889 (99\%) & \underline{ 14.21}  \\
            & TLP    & \cite{ZynqMar17}  & 11916 (85\%) & 15.1 \\ 
            & TLP-LLP &   & 13889 (99\%) & \underline{15.74} \\ 
            & PP    &  & 11916 (85\%) & 13.7 \\  
            & PP-TLP &  & 13861 (99\%) & \underline{14.09} \\
            \hline
        & BBLP & 15 000 & 14166 (94\%) & \underline{13.62} \\   
            & LLP  &  & 14722 (98\%) & \underline{14.7}  \\
                & TLP    &   & 14166 (94\%) & \underline{16.7} \\
            & TLP-LLP &   & 14471 (96\%) & \underline{16.9} \\ 
                & PP    &  & 14166 (94\%) & \underline{16.5} \\
        & PP-TLP &  & 14166 (94\%) & \underline{\textbf{18.31}} \\
                \hline
& BBLP & 30 000  & 14166 (\textcolor{applegreen}{47\%}) & 13.62 \\  
            & LLP & \textbf{Artix Z-7012S } & 29773 (\textcolor{Crimsonglory}{99\%}) & \underline{16.3}  \\ 
& TLP    & \cite{ZynqMar17}  & 14166 (\textcolor{applegreen}{47\%}) & 16.7  \\
            & TLP-LLP &   & 29773 (\textcolor{Crimsonglory}{99\%}) & \underline{18.24} \\ 
& PP    &  & 14166 (\textcolor{applegreen}{47\%}) & 16.5 \\
& PP-TLP    &  & 14166 (\textcolor{applegreen}{47\%}) & \textbf{18.31} \\
                \hline                  
  \end{tabular}
  }  
  \caption{Area Budget and Area Used for audio decoder.
  }
  \vspace{-0.2cm}
  \label{tab:decoder}
\end{table}

\subsection{Aladdin/gem5 and Catapult HLS}

\begin{figure}[t]
\centering
\vspace{-0.2cm}
\includegraphics[width=0.55\linewidth]{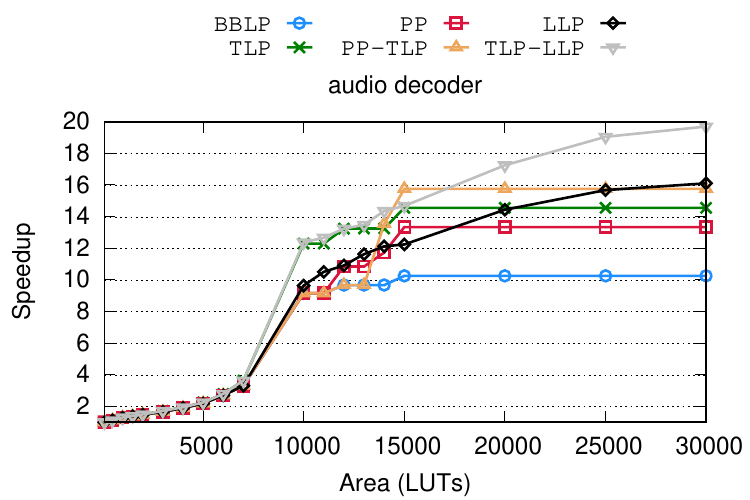}
\vspace{-0.3cm}
\caption{
Speedup obtained for audio decoder varying the area constraint using Aladdin \cite{ShaoJul14} for the HW acceleration parts and gem5 \cite{BinkertFeb11} for the SW-only implementation.
}
\vspace{-0.1cm}
\label{fig:audio_decoder_aladdin}
\end{figure}

To validate the selection of the HW/SW designs for every parallelism strategy explored and evaluated by our tool-chain, we use Aladdin \cite{ShaoJul14}, a HW accelerator simulator, and the gem5 \cite{BinkertFeb11} simulator. 
Aladdin was chosen as a faster, yet accurate, alternative to commercial HLS tools that offer latency and area results.
For \decoder, we gather the HW latency and area of the available candidates for acceleration with Aladdin, and their respective SW latency with gem5, as well as the run-time of the application as detailed in Section \ref{sec:setup}.\par

Figure \ref{fig:audio_decoder_aladdin} shows the speedup over increasing area budgets. For every area budget, the outputs of applying the parallelism strategies explored in this work match the ones generated by the Aladdin/gem5 simulations. This reinforces our expectation that our tool-chain selects the most promising designs with respect to performance and area usage.\par

As expected, speedup absolute values for \decoder\
 (Figures \ref{fig:audio_decoder} and \ref{fig:audio_decoder_aladdin})
differ. This is due to two factors: A) Our performance and area models are not based on cycle-accurate estimations, but aim to enable the selection of high-performance HW/SW choices automatically, and faster than performing demanding simulations or RTL synthesis. B) The characterization of latency for Aladdin is performed targeting 
OpenPDK 45nm technology, which is different to the characterization of our tool targeting a Zynq Programmable 
SoC.

To further evaluate our tool flow,
we designed accelerator prototypes using SystemC, guided by \toolname. To gather HW latency and area requirements, the accelerators were synthesized using Catapult HLS \cite{CatapultHLS}. The RTL was then synthesized, placed and routed by ASIC EDA tools using a commercial 12nm FinFET technology. The accelerators were clocked at 500MHz frequency and cycle-accurate Catapult simulations were used to measure the HW latency.\par

\begin{figure}[h]
\centering
\vspace{-0.5cm}
   \hspace{1.4cm}
\includegraphics[width=0.28\linewidth]{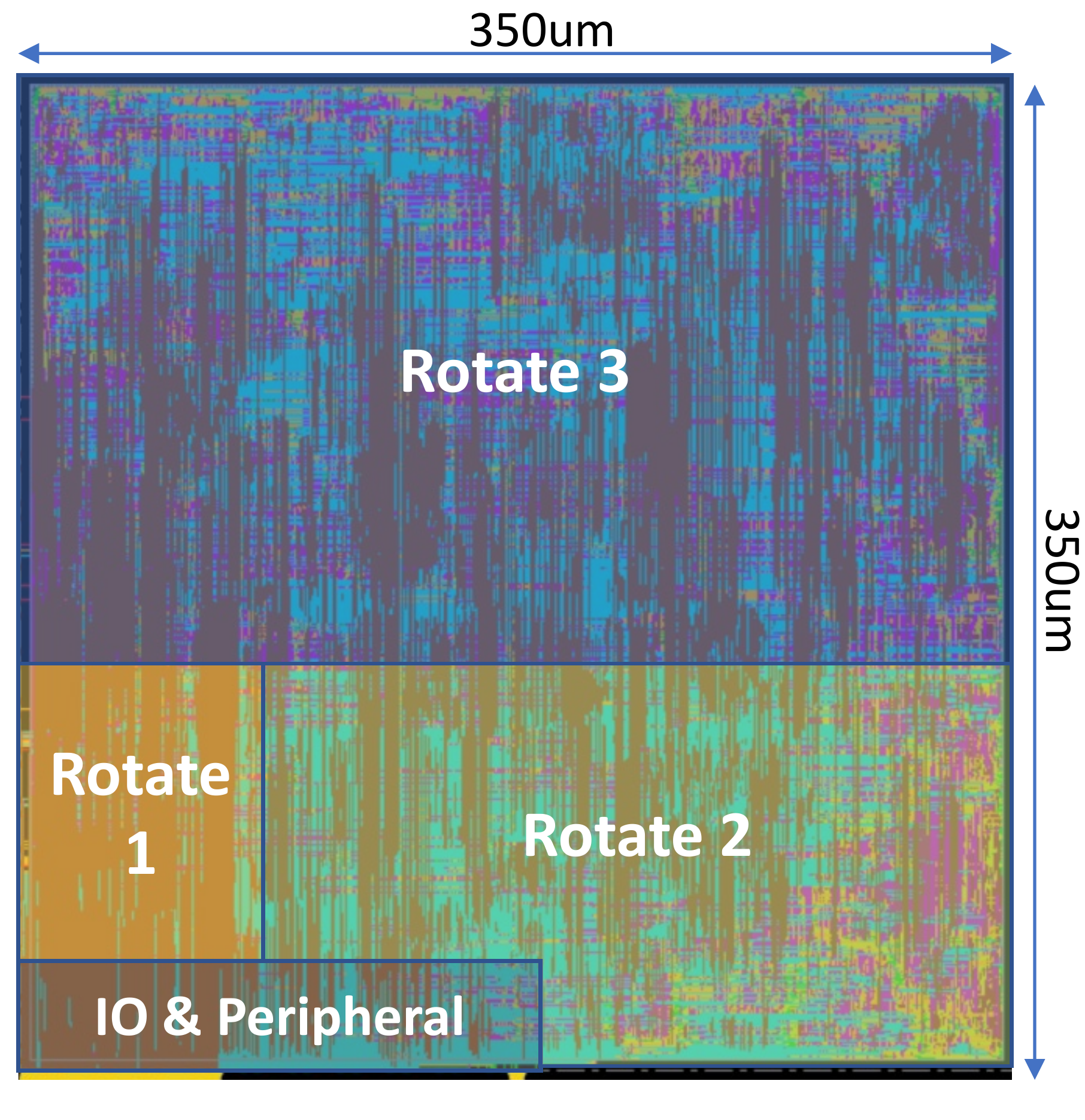}
\includegraphics[width=0.15\linewidth]{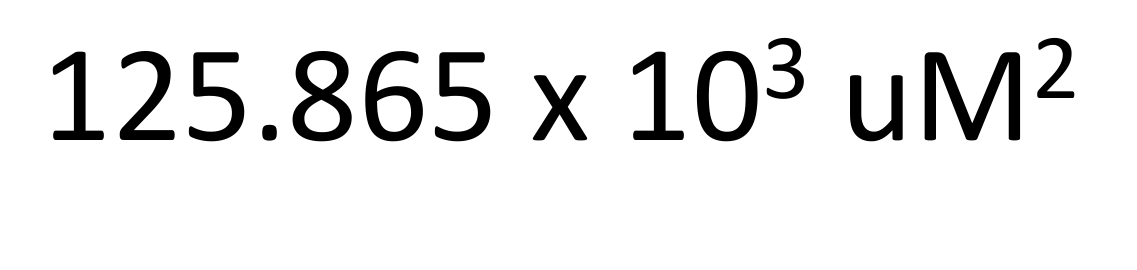}
\vspace{-0.2cm}
\caption{HLS design of audio decoder guided by \toolname.
}
\label{fig:layout}
\end{figure}

Table \ref{tab:catapult}
shows the HW latency comparison of Trireme (LLP, TLP, TLP-LLP) to AccelSeeker (BBLP).
For \encoder, LLP designs guided by \toolname\ achieve impressive performance gains at the expense of more HW resources.
In \decoder, LLP designs achieve smaller speedup and require the same or more resources compared to TLP-LLP. The latter can be up to six times faster compared to the respective \aseeker\ design for a large area budget (\textasciitilde$252 \times 10^3 uM^2$).
A medium area budget of \textasciitilde $126 \times 10^3 uM^2$ can yield significant speedup for TLP and TLP-LLP where accelerators Rotate 1-3 are operating in parallel.
Figure \ref{fig:layout} shows the physical layout of this design for \decoder.

  \begin{table}[h]
  \footnotesize
  \centering
  \resizebox{0.6\linewidth}{!}{

    \begin{tabular}{|l|c|c|c|} 
\hline
  \textbf{Benchmark}    & \textbf{Parallelism}    & \textbf{Area Used} & \textbf{Speedup vs.}   \\
  & \textbf{Version}  & ($uM^2$) & \textbf{ \soTa} \\ 
  & & & \textbf{\aseeker\ (BBLP)} \\ 
   \hline \hline
audio encoder & BBLP  & 3854 & 1 \\
\hline 
             & LLP    & 5415 & 2 \\
            \hline   
             & LLP    & 8578 & 4 \\
            \hline   
            & LLP  & 15072 & 8 \\
               \hline 
             & LLP    & 27491  & 16 \\
            \hline        
            \hline 
audio decoder & BBLP  & 92 738 & 1 \\
             & LLP    & 85 602 & 1.5 \\
            & TLP-LLP   & 85 602  & 2 \\  
            \hline
        & BBLP  & 125 865 & 1 \\   
            & LLP   & 171 385 & 2  \\
            & TLP    & \textbf{125 865} & \textbf{3} \\ 
            & TLP-LLP & \textbf{125 865} & \textbf{3} \\ 
            & TLP-LLP  & 251 641 & 6 \\ 
                \hline                  
  \end{tabular}
  } 
  \caption{
  \toolname\ vs. \aseeker\ \cite{ZacharopoulosNov19} by Catapult HLS 
  \cite{CatapultHLS}.
  }
  \label{tab:catapult}
  \end{table}

\subsection{Configurations of the Target Platform}

To gain better intuition on how different platform configurations affect potential speedup in HW accelerated systems, we apply a round of experiments varying the bandwidth of the data transfers to and from the HW accelerators (affecting memory latency), and the overhead of invoking them. Note that for Subsections \ref{sec:res_llp}, \ref{sec:res_llp_pp} and \ref{sec:res_llp_pp_tlp} we have been assuming a configuration of 1 GBps bandwidth and $1 \mu\text{s}$ overhead per accelerator invocation.\par

\begin{figure*}[h]
\centering
\hspace{-0.4cm}
\includegraphics[width=0.5\linewidth]{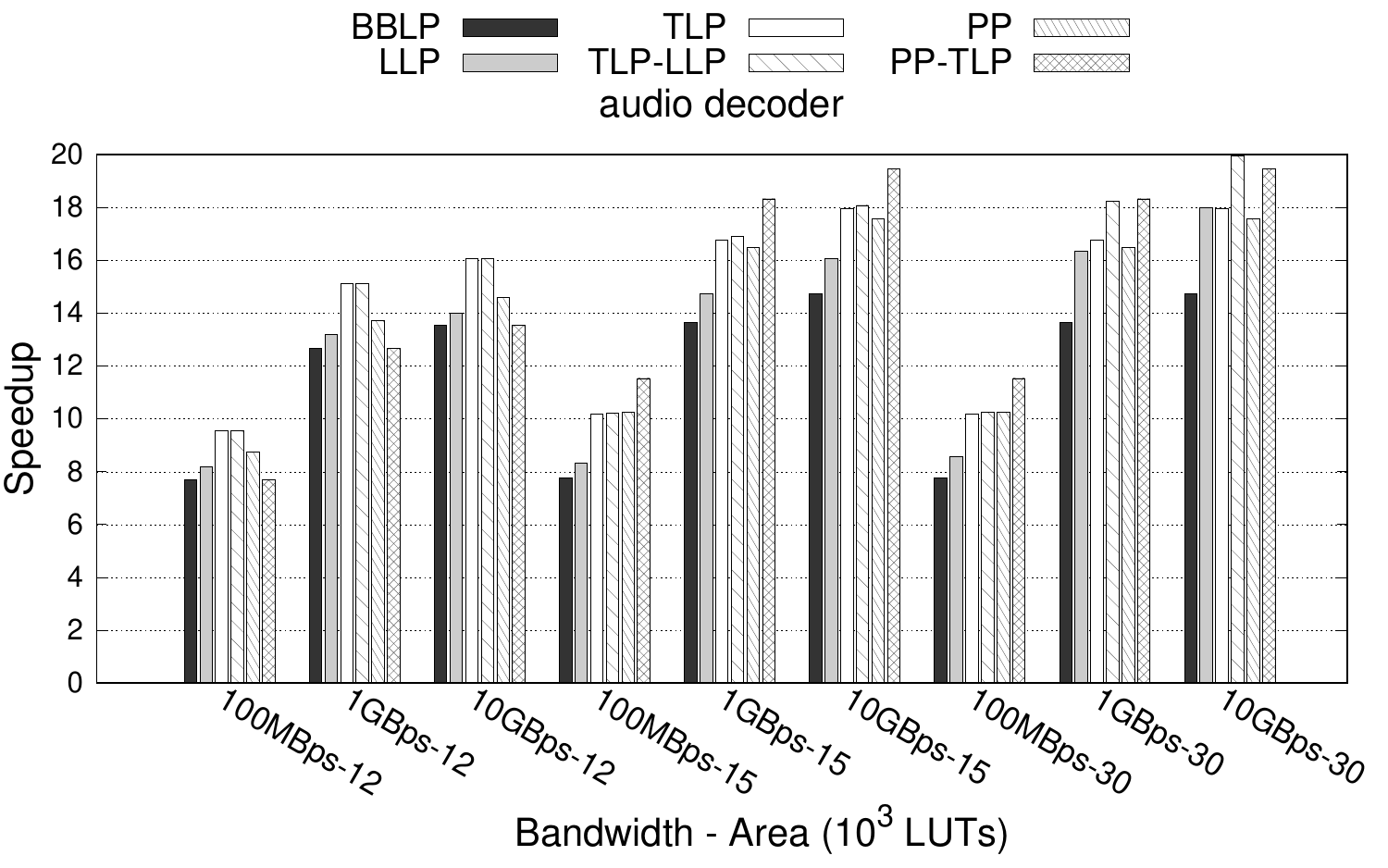}
\includegraphics[width=0.5\linewidth]{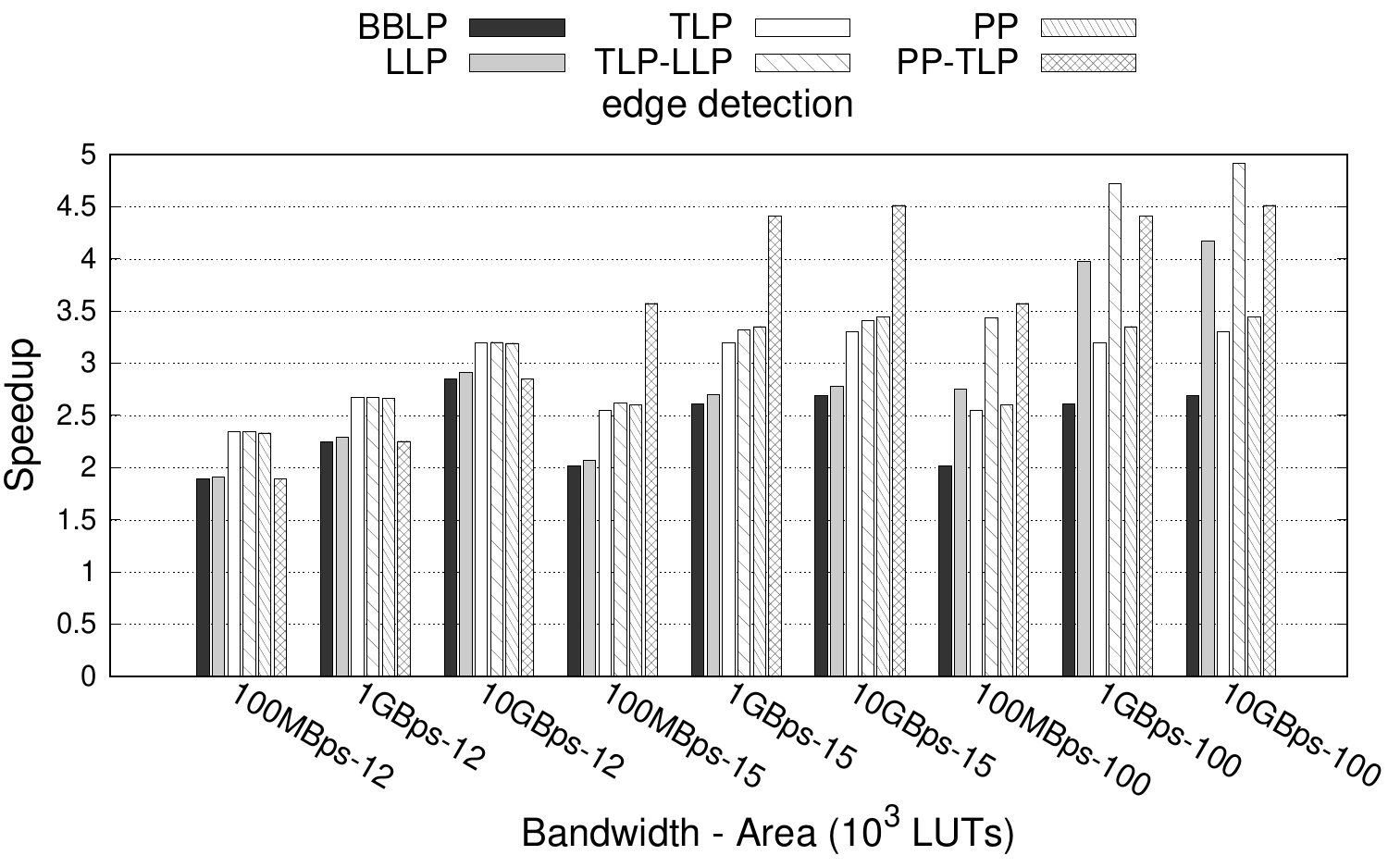}
\caption{Speedup of audio decoder (left) and edge detection (right), for increasing bandwidth and area. 
Baseline is SW-only.}
\label{fig:configs}
\end{figure*}

Figure \ref{fig:configs} (left) shows the \decoder\ speedup due to varying the bandwidth over 100 MBps, 1 GBps and 10 GBps, and the area budgets over 12, 15 and 30$ \times 10^3$ LUTs. We observe that low bandwidth (100 MBps), even when the area budget is increased, offers little speedup from BBLP, LLP, TLP, TLP-LLP and PP. 
This reveals the limitation of platforms where communication to memory can severely affect the speedup of a HW/SW design.\par

Overall, as expected, all parallelism strategies reach greater speedup when both bandwidth and area are increased. Nonetheless, LLP and TLP-LLP are favored, compared to the rest of the strategies,  when bandwidth is increased for a given area budget. This result is even more evident for \edge\ compared to \decoder, as seen in Figure \ref{fig:configs} (right), as it has more parallelizable loops than the latter. For the largest area budget of $100 \times 10^3$ LUTs we notice that the second and fourth bars increase vastly reaching 4.2$\times$ and 4.9$\times$ speedup respectively, as bandwidth increases, surpassing the previous better performing strategy (PP-TLP) for a smaller budget of $15 \times 10^3$. We can also notice this for \decoder\ for the largest area budget of $30 \times 10^3$ LUTs where TLP-LLP reaches the maximum speedup (20$\times$), compared to the rest of the parallelism approaches.\par

\section{Related Work}\label{sec:related}

We classify related research literature across five  dimensions, as shown in Table \ref{tab:taxonomy}. The types of parallelism supported by each piece of research vary from ILP within Basic Block boundaries \cite{ZacharopoulosApr19, ZacharopoulosNov19}, to \llv\ \cite{nardi2019hypermapper, koeplinger2018spatial,peruse, DurstFeldman2020}, \tlp\ \cite{ nguyen2016fcuda, margerm2018tapas} and Tensor level \cite{circt}.
Early DSE, one of the most important aspects of \toolname, is in many instances not supported by tools developed to expose and exploit parallelism in HW acceleration \cite{fcuda, peruse, margerm2018tapas, schardl2017tapir, circt}.

FCUDA~\cite{fcuda} is a source-to-source tool that translates CUDA code to FPGA accelerators, however offers no DSE or estimation of HW acceleration performance. On the other hand, Spatial \cite{koeplinger2018spatial} is an early DSE infrastructure that uses Hypermapper $2.0$~\cite{nardi2019hypermapper} in order to apply early DSE, however the parts to be accelerated need to be user-defined and high level languages are not supported as input. Aetherling \cite{DurstFeldman2020} applies early DSE as well and can be configured onto FPGAs, but it is restricted to loop level parallelism only and, like Spatial, does not support high level languages (C/C++).\par

Methodologies that combine static analysis and machine learning have been used in Peruse \cite{peruse} and in \cite{ZacharopoulosJul18} to predict the potential speedup of loop accelerators.
TAPAS~\cite{margerm2018tapas} is a tool-chain focusing on loop and task level parallelism by leveraging the TAPIR~\cite{schardl2017tapir} Parallel IR representation of the code. Although TAPIR is able to generate parallelism at arbitrary granularities, HPVM is able to expose nested parallelism which is leveraged by \toolname.\par
 
HeteroCL~\cite{lai2019heterocl}, developed within a Python-based domain specific language, performs early DSE and offers estimations on performance and area targeting FPGAs. It uses parallel processing pipelines and shifts towards tensor-related computations, used in Linear Algebra, Computer Vision and Machine Learning. Since HeteroCL is domain specific, it uses the domain expertise to trade accuracy for performance aggressively by reducing the bitwidth for key functional units. \par

High Level Synthesis (HLS) tools have improved substantially in
recent years \cite{MeeusSep12}. Commercial tools like 
Xilinx Vivado HLS \cite{VivadoHLSMar17} and Cadence Stratus HLS \cite{StratusHLSApr16}, and academic tools like Bambu \cite{PilatoMar12} and Legup \cite{CanisSep13b},
carry out the design of computation-heavy accelerators from application source code. They achieve performance on a par with that of hand-crafted implementations written in low level hardware description languages like VHDL and Verilog. But these HLS tools provide no DSE or early estimation of accelerator performance; hence, they are complementary to \toolname{} in an application-driven hardware-design workflow. \par

Tools that perform HW acceleration simulation and can be used for DSE such as Aladdin \cite{ShaoJul14}, gem5-aladdin \cite{gem5aladdin} and gem5-SALAM \cite{gem5salam}  
can achieve high cycle and power accuracy, comparable to that of commercial HLS tools. Furthermore optimizations, such as loop unrolling and loop pipelining, can be applied. However, a considerable amount of manual work is required and the simulation process is fairly time-consuming, even though significantly less than the time required by commercial HLS tools.
Finally, frameworks used for automatic binary parallelization \cite{zhou2019janus} and for automatic parallelization of non-numerical applications \cite{campanoni2014helix} by decoupling communication from computation, in order to avoid the overhead due to synchronization, have also been proposed.

\begin{table}[h]
  \Huge
  \vspace{0.7cm}
  \centering
  \resizebox{0.8\linewidth}{!}{
    \begin{tabular}{|l|c|c|c|c|c|c|c|c|} 
  \hline
  \textbf{Feature}    & \textbf{FCUDA}   &  \textbf{Spatial} & \textbf{Peruse} & \textbf{TAPAS} & \textbf{CIRCT} & \textbf{Aether} & \textbf{Accel} & \textbf{\toolname}  \\
  &  & \textbf{} & &   & & \textbf{ling} & \textbf{Seeker} &  \\ 
      &  &  & &   &  &  & & \\ 
    & \cite{fcuda} & \cite{koeplinger2018spatial, nardi2019hypermapper}  & \cite{peruse} &  \cite{margerm2018tapas} & \cite{circt}  & \cite{DurstFeldman2020} & \cite{ZacharopoulosNov19} &  \\ 
   \hline \hline
  Levels of    & Loop  & Loop & Loop &Loop & Tensor & Loop & Intra-BB  & Intra-BB\\ 
  Parallelism & Task  & Task  &  & Task   &   &  & ILP & Loop \\ 
     &   &   &  &    &   &  &  & Task \\ 
        &   &   &  &    &   &  &  & Pipeline \\ 
  \hline
  Early          & \redtick & \greentick & \redtick & \redtick & \redtick & \greentick & \greentick & \greentick \\ 
   DSE &  & &  &    &   &  &  & \\ 
  \hline
  Performance   & &  &  &  &  &  &  & \\
  Estimation   & \redtick & \greentick & \greentick & \redtick   & N/A &  \greentick & \greentick & \greentick \\ 
  \hline
  Automated  &  \greentick & \greentick & \greentick &  \greentick & \greentick & \greentick & \greentick & \greentick \\ \hline
  Configurations &  &  & &  &  &  &  & \\ 
  of Target   &  & & &   &  &  &  &\\ 
  SoCs & \redtick & 
  \redtick & 
  \redtick &  \redtick & N/A & \redtick & \redtick & \greentick\\ 
  \hline 
  \end{tabular}
  }  
  \caption{Taxonomy Table.}
  \label{tab:taxonomy}
\end{table}


\section{Conclusions}
Early DSE in modern applications, along with the extraction of critical information about parallelism, can be crucial to the outcome of a final HW/SW design and its respective performance on SoCs.
\toolname\ leverages information automatically retrieved by HPVM and applies it to accelerators automatically identified and evaluated by \aseeker. Using novel performance models, \toolname{} is able to thoroughly explore a variety of parallelism strategies and select the highest performing HW/SW design as output for area budgets of increasing size. We have explored multiple SoC configurations, varying the data transfer bandwidth between memory and accelerators, as well as accelerator invocation overhead. Application of \toolname\ to the XR domain yields substantial speedup gain with fixed resources when compared with
state-of-the-art tools (e.g., \aseeker\ \cite{ZacharopoulosNov19}) 
that do not consider \llp, \tlp{} and \pp.

\bibliographystyle{ACM-Reference-Format}
\bibliography{sample-acmsmall}

\end{document}